\documentclass[twoside]{article}
\usepackage{fleqn,espcrc2}

\usepackage{epsfig}
\newcommand{\AmS}{{\protect\the\textfont2
  A\kern-.1667em\lower.5ex\hbox{M}\kern-.125emS}}
\title{Difference between microscopic and effective overlaps in the
copper-oxide planes of high-Tc superconductors}

\author{Ladislav Jankovi\'c and D.K. Sunko
        \thanks{email: dks@phy.hr. Work supported by the Croatian Government
        under Project~$119\,204$.}\\[-1ex]\vskip\baselineskip
        Department of Physics, Faculty of Science, University
        of Zagreb,\\ Bijeni\v cka cesta 32, HR-10000 Zagreb, Croatia.}
       
\begin{document}

\begin{abstract} 

We investigate parametrizations required to reproduce observed Fermi surfaces
of overdoped and slightly underdoped LSCO in two different models. One is the
standard three-band saddle-point slave boson (SB). The other is a model of
CuO4 `molecules' randomly tiled in the plane, emphasizing the distinction
between the local overlap, always the full (bare) one, and the effective
overlap, connected with the reduced bandwidth due to the random tiling (RT).
We conclude this distinction is physically significant for underdoped, but
not for overdoped LSCO. This is consistent with the observed open Fermi
surfaces being due to correlation effects.

\vspace{1pc}
\end{abstract}

\maketitle

\section{Introduction}

The nature of the normal state of high-T$_c$ superconductors is still
an open question. Within ARPES, the discussion has initially
centered on line shapes~\cite{line}, but with improved resolution and material
preparation, has moved to pseudogap behavior~\cite{pseudo} and Fermi surface
evolution~\cite{lsco}.

The present work investigates, does the recently observed~\cite{lsco} change
of shape of the Fermi surface in La$_{2-x}$Sr$_x$CuO$_4$ with doping by
itself require invoking correlations. In particular: is there a sign that the
local electronic environment is different than the `effective' one-particle
picture which can reproduce these shapes.

\section{Model}

We start from a metallic state and three bands, and assume that the principal
effect of the infinite on-site repulsion is in charge correlations. This
regime was first described by the SB model at the saddle-point~\cite{klr}, by
renormalizing $t$. Physically, the local overlap is not always renormalized,
because when an electron of one spin is on a site, the other is may not be
near. This distinction between local and long-range behavior is lost at the
saddle-point.

The RT model~\cite{sb} keeps the distinction, by first hybridizing the
electron within a single CuO$_4$ `molecule' through the bare overlap, and
then producing an effective band by randomly tiling these molecules in the
plane. It satisfies the Pauli principle for the original electrons exactly. 
The price paid is that the one-particle formulation is achieved by assuming
that the down-spins do not hop at all. The fact that they do is simulated by
annealment. We cannot justify this microscopically; however, the assumption,
that the two spins see each other as static~\cite{gutz}, is implicit at the SB
saddle-point as well~\cite{kr}.

\section{Results}

To face measurements, we include the oxygen-oxygen overlap $t'$ in the RT
model in the simplest local approximation, just adding the $t'$ terms to
a CuO$_4$ `tile'. This underestimates the tendency of
$t'$ to rotate the Fermi surface, strenghtening our conclusion. The
implementation of the slave-boson model with $t'$~\cite{edo} follows the
literature.

\begin{figure}[t]
\begin{center}
\begin{picture}(6.9,5.75)
\put(0,-1){\psfig{
figure=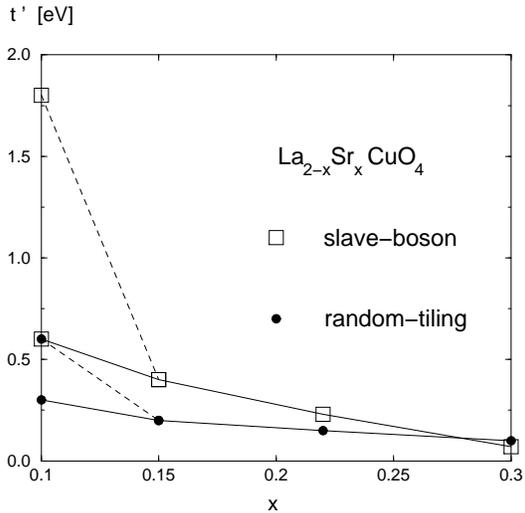
,width=6.9cm,height=6.75cm}}
\end{picture}
\end{center}
\caption{Full lines: values of $t'$ which fit Fermi surfaces within
error bars. Dotted lines: changes in $t'$ required for `perfect' fits.
For all points, $t=1$~eV, $\Delta_{pd}=3$~eV.
\label{dop}}
\end{figure}

In fig.\ 1, we compare the RT and SB input parameters needed to fit observed
Fermi surfaces~\cite{lsco}. Both models can fit all the data equally well at
all dopings. The fit was sensitive to $t'$, while less was gained by varying 
$t$ and the bare copper-oxygen splitting $\Delta_{pd}$, so they were kept
fixed.

Overdoped samples ($x>0.2$) require a rather small value of $t'\sim 0.1$~eV,
and the two models nearly agree. For the underdoped ($x=0.1$) sample, the
discrepancy is a factor of two (0.6~eV for SB, 0.3~eV for RT). It is
known~\cite{kl} that such, fairly large, values of $t'$ are required to
rotate the Fermi surface in the SB model.

At $x\ge 0.15$, the fits are `too good', passing near data points, not merely
within error bars. However, if the same `perfect' fits are required at
$x=0.1$, $t'$ needs to be much larger, increasing the discrepancy factor to
three.

\section{Discussion}

Fourier series being efficient in fitting smooth curves, the mere fact that a
dispersion with more than nearest neighbors fits a 2D Fermi surface is hardly
significant. A credible phenomenology needs to be both sensibly constrained,
and require little variation in the input parameters. Cases where it breaks
down should also be understood. The fits in this work were constrained by
modelling, with only the bare parameters put in. Also, only $t'$ varied with
the data, which amounts to `freezing' the lattice. Although technically no
more than fits within error bars are required, it is significant that the
data overtax both models at underdoping, as evident by drastically increased
$t'$ when pushed to overfit.

The RT model treats the data more sparsely, with a range in $t'$ (full lines)
of 0.2~eV, \emph{vs.}\ 0.55~eV for the SB model. This, and the quantitatively
different tendency to breakdown at $x=0.1$, leads us to conclude that the
distinction between local and effective overlap is physically relevant for
the underdoped and, probably, optimally doped samples, but not for $x>0.2$.
We find open Fermi surfaces at $x=0.1$ and $0.15$ as the result of
correlations, at a reasonably low bare $t'$. This corresponds to one of the
regimes where an extended vH singularity can be obtained in the copper
bonding band~\cite{mb}. The 0.2~eV variation of $t'$, still remaining in the
RT model, could be due to our `local' approximation in including it, lattice
evolution, and spin correlations~\cite{mae}.

We thank A.~Fujimori for sending us data~\cite{lsco}, E.~Tuti\v s for his SB
program~\cite{edo} and reading the manuscript, and S.~Bari\v si\'c for
discussions.

\end{document}